\documentstyle[twoside,fleqn,espcrc2,epsf]{article}

\def\Tr{\mbox{ Tr }}
\def\Re{\mbox{ Re }}
\def\Im{\mbox{ Im }}

\title{Dirac operator as a random matrix and the quenched limit of QCD
with chemical potential}

\author{M. A. Stephanov\address{Department of Physics,
				University of Illinois,
				1110 West Green Street,
				Urbana, IL 61801-3080        }%
\thanks{This work was supported by NSF-PHY92-00148.}
}

\begin{document}

\begin{abstract}
The behavior of quenched QCD at nonzero chemical potential $\mu$ has
been a long-standing puzzle.  An explicit solution is found using the
random matrix approach to chiral symmetry breaking. At nonzero $\mu$
the quenched QCD is not a simple $n\to0$ limit of a theory with $n$
quarks: a naive `replica trick' fails. A limit that leads to the
quenched QCD is that of a theory with $2n$ quarks: $n$ quarks with
original action and $n$ quarks with conjugate action.
\end{abstract}

\maketitle

\section{The puzzle}
The main result of the present study is a resolution of a puzzle which
existed since the middle of 80's and has been one of the major
obstacles towards our understanding of the behavior of QCD at finite
baryon density. The contradiction is simple and acute.  The quark
chemical potential $\mu$ in QCD shifts the energy of a given quantum
state by $3\mu$ times this state's net baryon number. A transition at
$T=0$ should occur when $\mu=\mu_c\approx m_p/3$: the vacuum state
without baryons is no longer the lowest energy state. It is believed
that the chiral symmetry is restored in the new phase.  Testing this
prediction by Monte Carlo simulation is difficult since the
determinant of the fermion matrix becomes complex at nonzero
$\mu$. Nevertheless one can use quenched approximation in this
case. Such Monte Carlo studies \cite{Ba86} give consistently the
critical value of $\mu_c\approx m_\pi/2$ --- a clearly unphysical
result, since the pion does not carry baryon number.

The present study is motivated by a desire to clarify the problem
using random matrix approach. 
On an exactly solvable model of chiral symmetry breaking
in QCD we demonstrate explicitly a general phenomenon related to the
nature of the quenched limit. Namely, one can view the quenched theory
as a somewhat abstract limit $n\to0$ of QCD with variable number $n$
of quark flavors. Then if the limit is smooth the quenched theory 
is a reasonable approximation to the
theory at finite $n$. In fact, this limit
is not always smooth and can be different from the quenched theory as,
e.g., in a theory of spin glasses.
We show that the quenched limit is not smooth in QCD at
finite $\mu$. Quenched theory is therefore not a good approximation to
full QCD at finite $\mu$. This resolves the old
contradiction. Moreover, we can describe a theory to which quenched
QCD {\em is} a good approximation.  It is a theory (suggested in
\cite{Go88}) in which each quark has a conjugate partner, essentially,
with opposite baryon number assignment.

\section{$\langle\bar\psi\psi\rangle$ and $\rho$}

We study analytical properties of the quark
condensate $\langle\bar\psi\psi\rangle$ --- the order parameter of the
chiral symmetry breaking --- as a function of the bare quark mass
$m$ which we allow to go into the complex plane: $m\equiv
z=x+iy$. Integrating out the quark degrees of freedom we can express
$\langle\bar\psi\psi\rangle$ through the average of the inverse Dirac
operator $D$:
$
\langle\bar\psi\psi\rangle = \langle \Tr (z - D)^{-1} \rangle 
\equiv G $.
In the infinite volume the value of the
$\langle\bar\psi\psi\rangle$ in the chiral limit $z\to0$ 
can be related to the density of small
eigenvalues $\rho(\lambda)$ by the Banks-Casher formula:
$\langle\bar\psi\psi\rangle = \pi \rho(0)$.

To generalize this well-known relation to our case we must consider
the density per unit {\em area} of the complex plane: $\rho(x,y)$. 
The relation between $\rho$ and $G$ is straightforward:
$G(x,y) = \int dx' dy' \rho(x',y') (z-z')^{-1}$. 
Viewing $\rho$ as a density of
charges and $\Re G$, $-\Im G$ as components of the electric field
strength $\vec G$ one easily inverts this relation:
\begin{equation}
\rho = {1\over 2\pi} \vec\nabla\vec G 
\equiv {1\over\pi} \Re {\partial G\over\partial z^*}.
\end{equation}
The main point here is that $\rho$ vanishes where $G$ depends
analytically on $z$, the bare quark mass. For example, when
$\mu=0$ all eigenvalues of $D$ lie on the imaginary axis. This means
that $G$ has a cut as on Fig. \ref{fig:n}a. The discontinuity across the
cut is the signature of the chiral symmetry breaking and is given
precisely by the Banks-Casher formula.  

\section{Random matrices}

Now we would like to calculate $G$ and $\rho$. 
Since we are interested in the density of {\em small}
eigenvalues of the Dirac operator one can expect that some simplification
can be made. Indeed, it is by now well understood that these small
eigenvalues are related to the zero modes of the Dirac operator in the
instanton field background. The behavior of these small eigenvalues
is rather universal and can be described by a simple random
matrix model \cite{ShVe93}.
We choose chiral representation of the Dirac gamma-matrices. Then the
matrix of the Dirac operator in a certain basis 
has the block-diagonal form:
\begin{equation}\label{D}
D=\left(\begin{array}{cc}
0 & iX \\ iX^\dagger & 0
\end{array}\right)
+\left(\begin{array}{cc}
0 & \mu \\ \mu & 0
\end{array}\right).
\end{equation}
Nonzero entries $X$ are approximated by a complex random $N\times N$
matrix with a Gaussian distribution: 
$P(X) \sim \exp\left\{-N \Tr XX^\dagger\right\}$.  To adapt
the model to our case we add a chemical potential:  $\mu\gamma_0$.

To find $G$ one calculates the quenched free energy:
\begin{equation}\label{V}
V = - \langle\ln\det(z-D)\rangle,
\end{equation}
and differentiates it with respect to the bare quark mass $z$:
$G=-\partial V/\partial z$. Then $\rho$ is given by:
\begin{equation}\label{rhoV}
\rho = - {1\over\pi} \Re {\partial^2 V\over\partial z\partial z^*}.
\end{equation}

A standard way to deal with the logarithm inside the averaging in
(\ref{V}) is the `replica trick'. One does
the calculation for arbitrary number of fermion species $n$:
\begin{equation}\label{Vn}
V_n = - {1\over n}\ln\langle{\det}^n(z-D)\rangle,
\end{equation}
and then takes $n\to 0$ hoping that the limit is smooth. Doing this
calculation we find that, first of all, $G$ is a holomorphic function
of $z$. It is given by a solution of a cubic equation:
$G((z+G)^2-\mu^2) - (z+G) =0$. It is analytic except for branch points,
where 2 solutions coincide, connected by cuts. The cuts follow lines
where the value of $\Re [G^2 - \ln((z+G)^2-\mu^2)]$ on two solutions
coincide. The physical sheet is determined by an asymptotic condition
at $z\to\infty$: $G\to1/z$, from the normalization of $\rho$.

\begin{figure}[t]
\vskip 3mm
\centerline{\epsfxsize 65mm\epsfbox{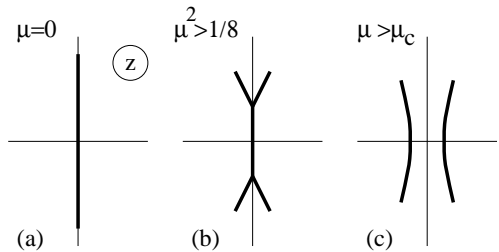}}
\vskip -6mm
\caption[]{The evolution of cuts of the function 
$G(z)\equiv\langle\bar\psi\psi\rangle$ with increasing $\mu$ 
from (\ref{Vn}).}
\label{fig:n}
\vskip -4mm
\end{figure}

The singularities of $G$ are shown in Fig. \ref{fig:n}. For
$\mu^2\leq1/8$ the cut is on the imaginary axis. At $\mu^2>1/8$ the
branch points bifurcate and the cuts go into the complex plane. Above
a certain value of $\mu_c^2=0.278\ldots$ the cut no longer goes
through $z=0$ --- there is no discontinuity across the imaginary axis,
i.e., the chiral symmetry is restored.

\section{$n\to0$ and quenched theory}

\begin{figure}
\vskip 3mm
\mbox{
\hskip -12mm
\epsfxsize 50mm \epsfbox{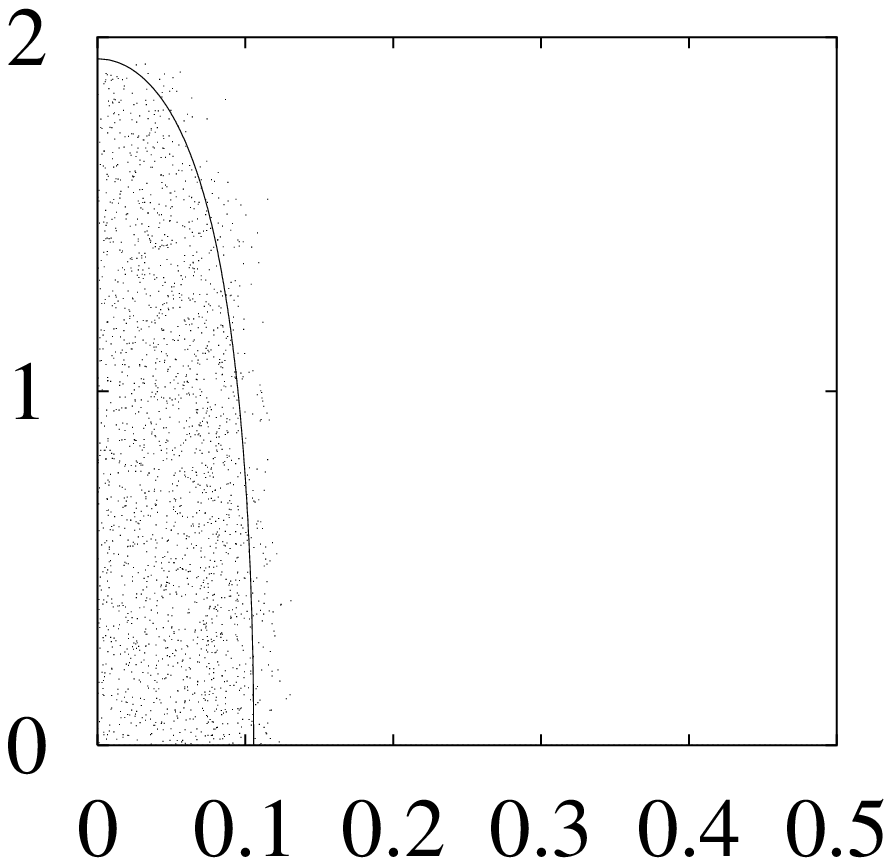} 
\hskip -11mm
\epsfxsize 50mm \epsfbox{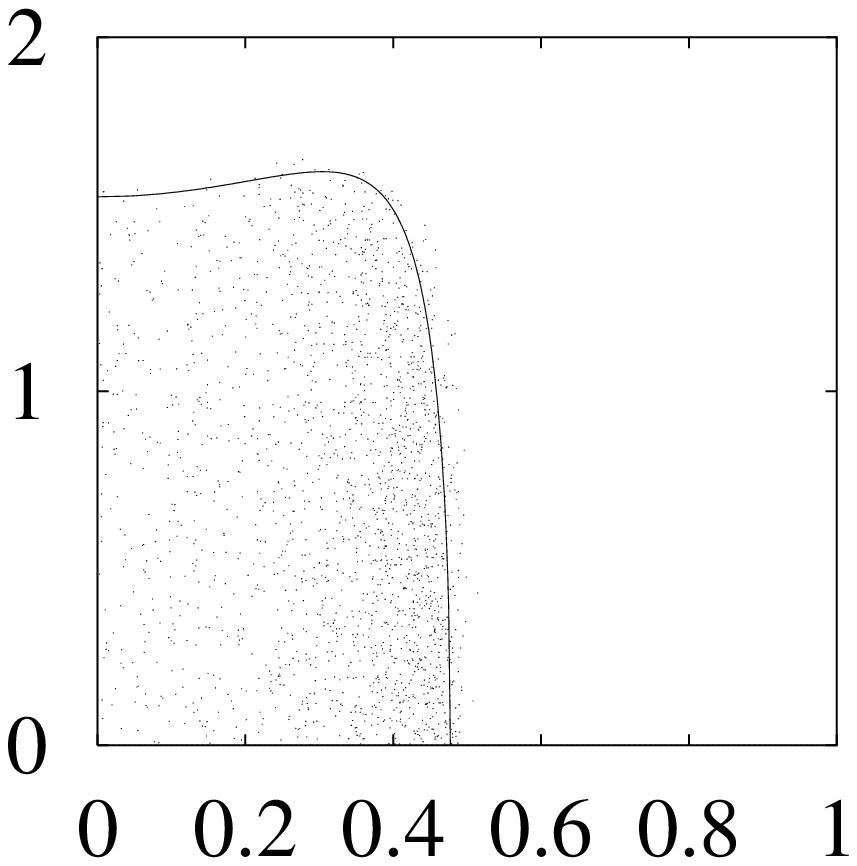}
}
\mbox{
\hskip 16mm (a) \hskip 32mm (b)
}\hfill
\vskip -6mm
\caption[]{Eigenvalues of 20 random $100\times100$ matrices on the
complex plane $z$ at two values of $\mu^2$: (a) 0.06, (b) 0.40.
The line is the boundary of the blob calculated analytically from
(\ref{Vnn}).}
\label{fig:q}
\vskip -4mm
\end{figure}

One can now compare this replica trick calculation to the actual
distribution of eigenvalues of the random matrix $D$ in the quenched
theory. It is obtained numerically, by plotting the eigenvalues for an
ensemble of random matrices (\ref{D}) on Fig. \ref{fig:q}.  First of
all, the eigenvalues do not stay on the imaginary axis for any nonzero
$\mu$! They spread in a blob of the $x$-width growing $\sim\mu^2$ at
small $\mu$. Exactly the same behavior is observed in the quenched QCD
calculations \cite{Ba86} and leads to the puzzling conclusion that
$\mu_c\approx m_\pi/2$ and vanishes in the chiral limit $m=0$ because
there is no discontinuity in $G$ across the imaginary axis when
$\mu>0$.

Now, however, we are able to find the explicit solution of the model and
see the nature of this quenched blob. First of all, we must
learn that the limit $n\to0$ of QCD with $n$ quarks is not smooth and
the quenched theory is not a good approximation to the full
theory at $\mu>0$. 
Second, we want to know what theory has quenched
QCD as its $n\to0$ limit. We find that it is a theory where each 
quark has a partner with the conjugate Dirac operator. The free energy
reads:
\begin{equation}\label{Vnn}
V_{n,n} = - {1\over n} \ln\langle {\det}^n (z-D)(z^*-D^\dagger) \rangle.
\end{equation}
This statement can be made rigorous even beyond the random matrix
approximation using equations (\ref{rhoV}) and:
$\delta^2(\vec z - \vec\lambda_i) =  - (1/4\pi) 
\Delta\ln|z-\lambda_i|^2.
$

Naively, when $n\to0$ the conjugate quarks decouple: $V_{n,n}\to V(z) +
V^*(z^*)$, and $G=-\partial V/\partial z$ is holomorphic. In fact, the
quarks $\psi$ and conjugate quarks $\chi$ mix for some values of
parameters $z$ and $\mu$, i.e., precisely inside the blob.  A
condensate $\langle\bar\chi\psi\rangle$ develops nonzero vacuum
expectation value.  This condensate carries nonzero baryon number. It
was indeed observed in a Monte Carlo simulation of the $SU(2)$ theory
{\em with} quarks \cite{Da86}, which are self-conjugate in that
case. We learn now that similar phenomenon occurs in the quenched
theory. Calculating (\ref{Vnn})
we find  \cite{St96} for the density of eigenvalues inside 
of the blob:
\begin{equation}\label{rho}
\rho = {1\over4\pi}\left( {{x^2+\mu^2\over(\mu^2-x^2)^2} - 1}\right),
\end{equation}
which one can check is in perfect agreement with the numerical data
Fig. \ref{fig:q}.

\section{Conclusions}

Using a simple random matrix model of chiral symmetry breaking we
demonstrate a general phenomenon that at {\em nonzero} $\mu$ the limit
$n\to0$ of QCD with $n$ quarks is not smooth and does not coincide
with the quenched theory. On the other hand, the quenched theory is a
smooth limit of a theory where each quark has a conjugate
partner. Another way of putting this is to notice that the difference
between such a theory and QCD with $2n$ quarks is that the phase of the
fermion determinant is omitted in the former. We learn that
the phase of the determinant is extremely important for a simulation
of full QCD at finite $\mu$. 
Without this phase the result $\mu_c=0$ is natural, as happens in
quenched QCD. This is due to the formation of the baryonic condensate,
which breaks the (replica) chiral symmetry of quarks and
conjugate quarks. The Goldstone boson associated with this breaking
is the so-called baryonic pion --- a bound state of a quark and a
conjugate antiquark. The presence of this particle explains the
result $\mu_c=m_\pi/2$.

\end{document}